\documentclass[a4paper]{article}
\usepackage{amsmath,amstext,amsgen,amsbsy,amsopn,amsfonts,amssymb}
\usepackage{easybmat}
\usepackage{graphics}
\usepackage[pdftex]{graphicx}
\usepackage{epsfig}
\usepackage{amsthm}
\usepackage{bm}
\usepackage{algorithm}
\usepackage{algorithmic}

\usepackage{wrapfig}
\usepackage{esvect}
\usepackage{multirow}
\usepackage{array}
\usepackage{booktabs}
\usepackage{xcolor}
\usepackage{hyperref}

\providecommand{\keywords}[1]{\textbf{\textit{Index terms---}} #1}
\usepackage{arydshln,booktabs, caption}

\begin{document}
\large
\title{\textbf{Towards Understanding the COVID-19 Case Fatality Rate}}
\author{
Donghui Yan$^{\dag}$, Aiyou Chen, Buqing Yang$^{\P}$
\vspace{0.1in}\\
\vspace{0.1in}\\
$^\dag$Department of Mathematics and Program in Data Science\\
University of Massachusetts, Dartmouth, MA 02747\\[0.04in]
$^\P$Department of Actuarial Science,\\ Shanghai University of Finance and Economics, China 
}

\date{\today}
\maketitle \normalsize
\begin{abstract}
\noindent
An important parameter for COVID-19 is the case fatality rate (CFR). It has been applied to wide applications, including the 
measure of the severity of the infection, the estimation of the number of infected cases, risk assessment etc. However, there 
remains a lack of understanding on several aspects of CFR, including population factors that are important to CFR, the apparent discrepancy 
of CFRs in different countries, and how the age effect comes into play. We analyze the CFRs at two different time snapshots, 
July 6 and Dec 28, with one during the first wave and the other a second wave of the COVID-19 pandemic. We consider two 
important population covariates, age and GDP as a proxy for the quality and abundance of public health. Extensive exploratory 
data analysis leads to some interesting findings. First, there is a clear exponential age effect among different age groups, and, 
more importantly, the exponential index is almost invariant across countries and time in the pandemic. Second, the roles played 
by the age and GDP are a little surprising: during the first wave, age is a more significant factor than GDP, while their roles have 
switched during the second wave of the pandemic, which may be partially explained by the delay in time for the quality and abundance 
of public health and medical research to factor in.
\end{abstract}
\section{Introduction}
\label{section:Intro}
\begin{figure}[ht]
\centering
\begin{center}
\hspace{0cm}
\includegraphics[scale=0.5,clip,angle=-90]{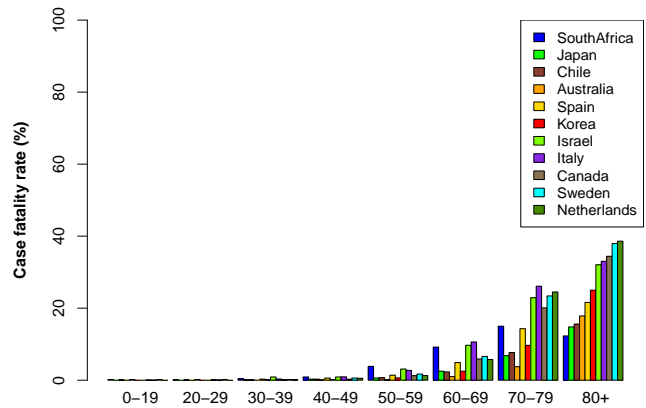}
\end{center}
\abovecaptionskip=-20pt
\caption{\it CFR by age groups for selected countries (as of July 6, 2020).  }  
\label{figure:cfrAgeCountries}
\end{figure}

The COVID-19 pandemic has quickly reached a global scale, with a total confirmed cases of 96.24 million and 
death toll at 2.06 million as of Jan 18, 2021. An important parameter for COVID-19 is the case fatality rate (CFR), which 
is {\it defined as the ratio of the death toll and the number of infected cases}. The primary use of CFR is as a quantitative metric for
the severity or lethality of the COVID-19 infection. It can be used as a reference in comparison to known infectious diseases 
such as the severe acute respiratory syndrome (SARS) or Ebola etc. An important application of CFR is to estimate the number 
of infected cases \cite{GuptaShankar2020,JagodnikLachmann2020} through the death tolls, as it is commonly believed 
that the death toll is a relatively reliable quantity.
It is also used as a proxy for risk assessment \cite{Schroder2020}. In order to apply the CFR properly, it is important to 
understand factors that contribute to CFR. While it is clear that the mortality of COVID-19 is closely related to the health 
status or pre-existing conditions of an individual, these are not suitable to understand CFR at the population level, for 
example at the scale of a country. COVID-19 death is often mixed with various other diseases related to the 
lung or cardiovascular diseases etc for an individual, which makes it challenging to characterize CFR at the population level. 
We need to understand CFR in terms of population parameters or covariates if we wish to understand the difference in CFR 
across different countries.
\\
\\
The population parameter we are primarily interested in is the age. It has been acknowledged there is a strong age 
effect in the mortality among COVID-19 cases---while the CFR for the seniors is high, 
it would be very low for young people especially those below 30 years old. This exponential disparity is illustrated 
in Figure~\ref{figure:cfrAgeCountries} which shows the CFR by age groups for a number of countries; the 
countries are selected primarily due to the availability of the data and turn out to distribute fairly evenly over the world.
It can be seen that, the CFR for people younger than 30 is almost 0 while increasing very rapidly among those older than 60. 
Though differing in details, this pattern is fairly consistent for all countries shown in the figure which are from different part of 
the world. However, as a matter of fact, countries in the world differ significantly in terms of their age profile. For example, many 
countries in Africa have a median age of around 20, while a significant portion of European countries have a median age over 40.  
We expect that the CFR for a young population be smaller than a population where senior people dominate. 
If one can clarify the age effect in CFR, that will help understand potential discrepancy caused by differences in age structures
across countries for comparing their CFRs, to assess how well a particular country or region (termed broadly as country from now 
on for simplicity of description) is doing in {\it controlling} the CFR,
or statistical inference of one country using CFR related information from another country. 
\\
\\
Other relevant population parameters include the quality and abundance of medical service or public health for a population, public policies, etc. 
The mortality of COVID-19 has been observed to be related to factors on the quality and abundance of health care and 
medical facilities, such as the number and capacity of hospitals and patient beds, testing coverage and accuracy, the quantity 
and quality of personal protection equipments, the experience of health workers and level of medical research on infectious disease 
etc. It is often challenging to quantify or
to access related data in many countries, and we will use the gross domestic production (GDP) per capita as a proxy for simplicity.   
\\
\\
We will carry out exploratory data analysis to investigate the role by age and GDP in CFR at the country level. We will start by 
considering the age effect, and then extend the analysis by including GDP.
The remainder of this paper is organized as follows. In Section~\ref{section:approach}, we will describe the methods. This is
followed by a presentation of data collection in Section \ref{sec:data} and the  results in Section~\ref{section:exp}. Section~~\ref{section:conclusion} concludes the paper. 

\section{Method}
\label{section:approach}
\begin{figure}[h]
\centering
\begin{center}
\hspace{0cm}
\vspace{-10pt}
\includegraphics[scale=0.5,clip,angle=-90]{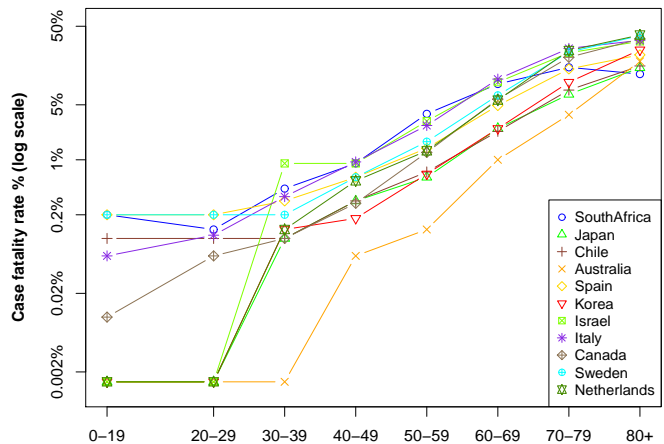}
\end{center}
\abovecaptionskip=-10pt
\caption{\it Log-scaled CFR by age groups for selected countries as of July 6, 2020.  }  
\label{figure:cfrAgeCountriesLog}
\end{figure}

The observed CFR for a given population can be very noisy. For example, the death toll may be affected by the use of potentially different definitions in counting 
mortality, the difficulty in determining the exact cause of death when COVID-19 is mixed with other chronic diseases, as well as missing counts or inflation in the reported case mortality, etc \cite{JagodnikLachmann2020}. Furthermore, the number of infected cases may be systematically under-counted since it is limited to patients who have access to testing. We analyze observed CFRs by fitting regression models which absorb all the noise into the error term. The goal is not to recover the underlying true CFR, but to unravel how age and GDP attribute to CFR across countries and over time.
Our method is partially motivated by the observation made in Figure~\ref{figure:cfrAgeCountries}, which tells that
at a crude level and in terms of the overall age trend, COVID-19 acts roughly similarly across different populations.
The major population covariates under consideration are age and GDP.
\\
\\
The regression models can be expressed as  $$Y_i=f(X_i, \theta) + \epsilon_i$$ where $X_i$ and $Y_i$ stand for the 
population covariates and the observed CFR for the i\textsuperscript{th} population, for $i=1, ..., n$, $\theta$ is the parameter
shared by all countries under consideration, and $\epsilon_i$ is used to model the noise 
in the observed CFR. Assume that $Y_i$'s are independent conditional on $X_i$. 
To be specific, we consider simple linear regression with $f(X, \theta)=\theta^T X$, which is powerful to discover strong main effects especially when the sample size is small.
\\
\\
Instead of using
the CFR directly, we use the log-scale, since the CFR appears to increase exponentially
with the age as evident from Figure~\ref{figure:cfrAgeCountries}. More directly, by visualizing the CFR in the log-scale as shown in Figure~\ref{figure:cfrAgeCountriesLog},
we see an almost linear increase (except for the age groups below 30) of the log-scaled CFR with the age. 
To better appreciate the magnitude of actual values of CFR for different age groups, we show as an example in Table~\ref{table:cfrCanada} 
the CFR by age groups in Canada. Alternatively, 
one may consider the {\it Logit} transform, that is, convert CFR to $\log(CFR/(1-CFR))$. As the CFR's are typically quite 
small, it is similar to the $\log$ transform. Though different in details, the overall linear pattern is fairly consistent 
across different countries. 

\begin{table}[h]
\begin{center}
\begin{tabular}{c||c|c|c|c|c|c|c|c}
\hline
Age    & 0-19         &20-29     &30-39       &40-49     &50-59          &60-69     &70-79   &  80+   \\
\hline
CFR              & 0.01\%     &0.06\%   &0.10\%     &0.28\%     &1.24\%       &5.90\%    &20.10\%    &34.42\%\\
\hline
\end{tabular}
\end{center}
\caption{\it CFR by age groups in Canada.} \label{table:cfrCanada}
\end{table}

\section{Data}\label{sec:data}

The data we use in our analysis includes the following. The number of reported cases and the death toll are retrieved 
from the Worldometer \cite{covidWorldOmeters}, which we use to calculate the observed CFR for individual countries in the world. 
The median age is taken from Wikipedia \cite{WikiWorldMedianAge}. The detailed age profile, i.e., percent by age groups, 
for countries is obtained
from the United Nations web \cite{UNPopulation}. The GDP per capita data is also taken from the Worldometer \cite{covidWorldOmeters}. 
Our initial analysis was carried out in the summer of 2020 using COVID-19 case data as of July 6, 2020. However, the pandemic 
had continued and deteriorated during the second half of the year. We were curious how that might impact our results. 
So we collect another snapshot of data, i.e., data sets as of Dec 28, 2020, also from the Worldometer.

\section{Results}
\label{section:exp}
In this section, we report results from the analysis on data collected at July 6, 2020 and Dec 28, 2020, respectively. 
We then make a comparison on these analysis, and report some interesting, maybe a little surprising, findings. 

\subsection{Analysis on data as of July 6, 2020}
\label{section:analysis0706}
\begin{figure}[h]
\centering
\begin{center}
\hspace{0cm}
\includegraphics[scale=0.5,clip,angle=-90]{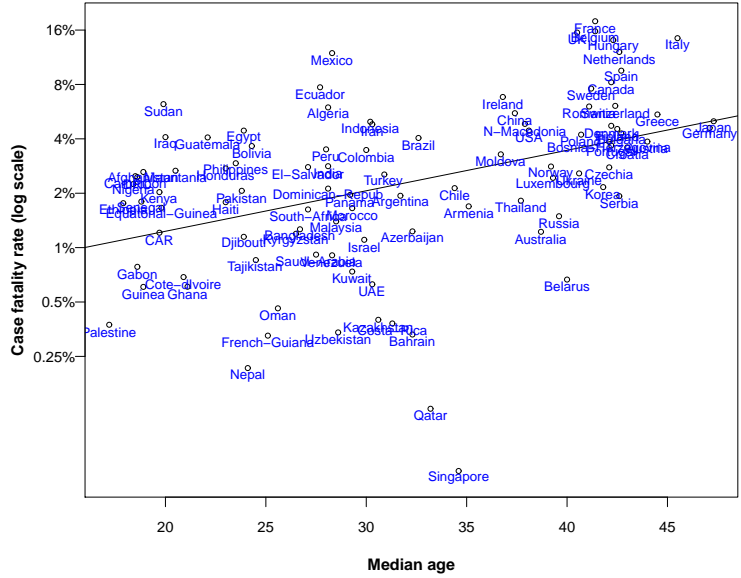}
\end{center}
\abovecaptionskip=-5pt
\caption{\it Scatter plot of CFR by median ages for individual countries as of Jul 6, 2020. The solid line is the regression line. }  
\label{figure:cfrAge}
\end{figure}

As of July 6, 2020, the observed CFR w.r.t. the median age for different countries is shown in Figure~\ref{figure:cfrAge}. 
There appears to be an overall increasing trend of CFR with the median age in the population. We start by considering the following 
simple linear model
\begin{equation}
\label{eq:lmAge0706}
\log(CFR) = \beta_0 + \beta_1 \cdot X + \epsilon,
\end{equation}
where $X$ is the median age of a population, and we term this as model I.
In carrying out linear regression model fitting, we exclude countries with less than reported 3000 cases as the CFR for 
such populations would be very noisy. This leaves us a total of 99 observations (i.e., countries) for linear regression; their total 
number of reported cases is 11,471,724 with a total death toll of 534,347. The fitted model parameters are
\begin{equation*}
\beta_0=-5.42877, ~ \beta_1=0.05160,
\end{equation*}
with a reported $R^2$ at 0.1726 (adjusted 0.164), and a p-value of $1.91\times 10^{-5}$ on F-test. All the coefficients are statistically significant 
with a p-value less than $1.91\times10^{-5}$. The fitted regression line is added as the solid line in Figure~\ref{figure:cfrAge}. 
As expected, the estimated CFR increases with the age of a population. 
Observed CFR in many countries indeed follow
this trend. 
\\
\\
With model \eqref{eq:lmAge0706}, we can estimate CFR for individual countries. 
For example, the CFR for the USA, India, China and Korea are
estimated as 3.13\%, 1.87\%, 3.02\% and 3.19\%, close to estimates at 2.85\% given by \cite{YanXuWang2020},
2.20\% by \cite{PhilipRaySubramanian2020}, 2.30\% by \cite{KugelgenGreseleSchkolpf2020}, 2.36\% by
\cite{ShimMizumotoChoi2020}, respectively. The worldwide CFR is estimated to be 2.76\%, close to the WHO published 3.40\%
as of Mar 2020; in contrast, a direct calculation from the reported cases and death toll would give 4.66\%. A country that stands out is Singapore which has 
extremely low observed CFR, given its above average median population age. We attribute this to the small size of this 
country and the painstaking efforts dedicated by its government in combating the pandemic. 
\begin{figure}[h]
\centering
\begin{center}
\hspace{0cm}
\vspace{-10pt}
\includegraphics[scale=0.56,clip,angle=-90]{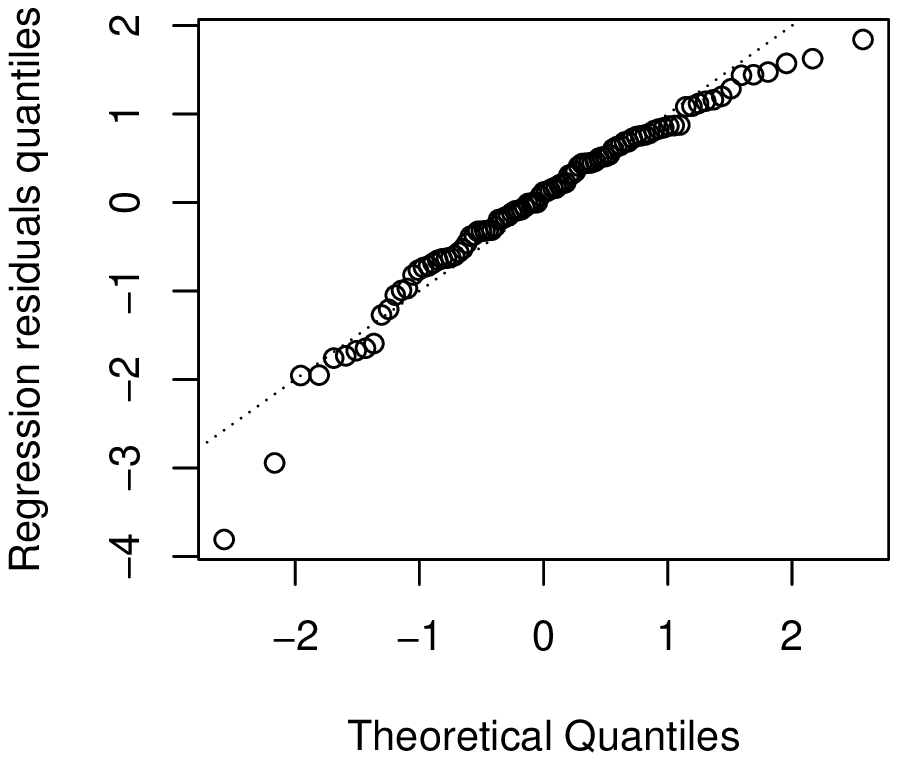}
\includegraphics[scale=0.56,clip,angle=-90]{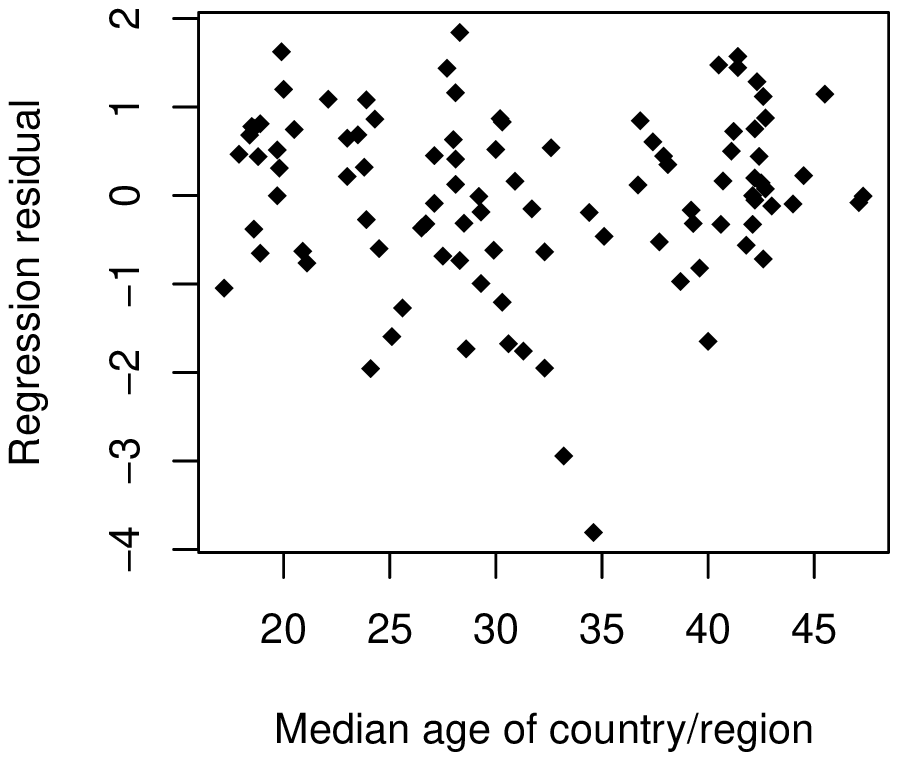}
\end{center}
\abovecaptionskip=-5pt
\caption{\it Regression diagnostics plots under Model I. The left and right panel are the QQ-norm plot of regression residuals and 
the residual plot, respectively. The dashed line in the QQ-plot is the y=x line. }  
\label{figure:regDiag}
\end{figure}
\\
In the linear regression analysis, we make two assumptions.
These include the assumption of conditional normality and that of the constant variance (i.e., identical $\sigma^2$ for different 
ages in the conditional normal density). To validate these assumptions, we carry out some regression diagnostic analysis \cite{Rice1995}. 
Figure~\ref{figure:regDiag} visualizes our results. The QQ-norm plot shows that, approximately, 
the regression residuals follow a normal distribution. We further perform a Kolmogorov-Smirnov test \cite{ChakravartiLahaRoy1967} 
of the regression residuals against a standard normal, which supports normality at a p-value of 0.374. 
Next, we look at the constant variance assumption. The residual plot shows that the regression 
residuals have a roughly constant spreadout over the range of median ages. The Cook-Weisberg's constant 
variance test \cite{CookWeisberg1983} gives p-value 0.8909, which suggests the compatibility of the data to homoscedasticity.
\\
\\
We can extend the above analysis by adding the GDP covariate, and term it as Model III. We code the GDP as 1 if it is smaller 
than \$10,000 per capita and 2 otherwise; the cutoff value of \$10,000 is close to that (i.e., \$12,000) used in determining if a 
country is a developing or developed country (indeed a cutoff value anywhere between \$8,000 and \$15,000 makes very little 
difference in our model). This yields the following fitted model parameters
\begin{equation*}
\beta_0=-5.255, ~ \beta_1=0.07140, ~\beta_2=-0.55369,
\end{equation*}
with a reported $R^2$ at 0.2132 (adjusted 0.1968), and a p-value of $1.006\times 10^{-5}$ on F-test. Using the original GDP 
value would lead to a slightly inferior model fit (with $R^2$ at 0.1851). The coefficient for the
age is statistically significant with a p-value less than $2.81\times10^{-6}$, but that for the GDP is not as significant with a p-value 
of 0.0284. 

\subsection{Analysis on data as of Dec 28, 2020}
\label{section:analysis1228}
\begin{figure}[hbtp]
\centering
\begin{center}
\includegraphics[scale=0.85,clip,angle=-180]{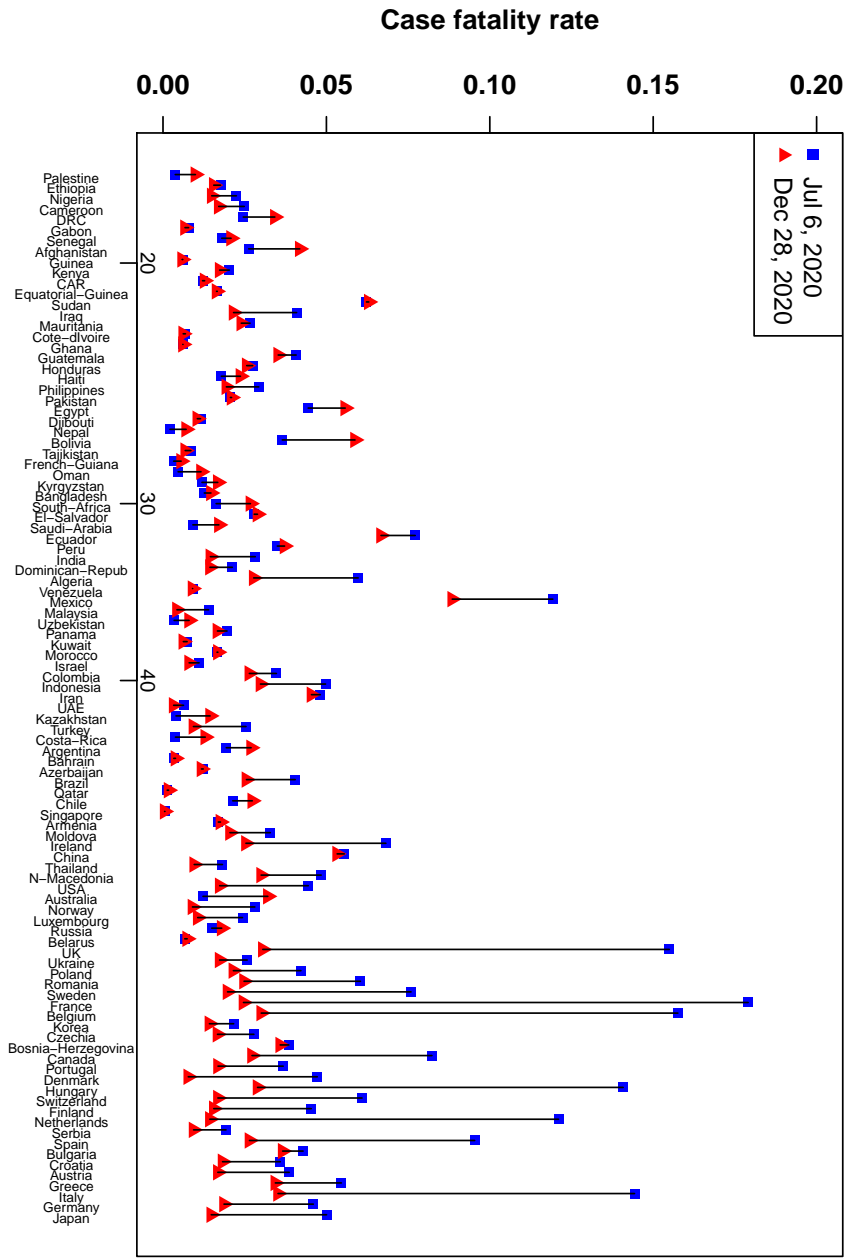}
\end{center}
\caption{\it Changes in CFR from July 6, 2020 to Dec 28, 2020. The countries are sorted by median ages in an increasing order
from left to right in the figure. The numbers on the x-axis are the median age. }  
\label{figure:cfrCompare}
\end{figure}

\begin{figure}[h]
\centering
\begin{center}
\hspace{0cm}
\includegraphics[scale=0.5,clip,angle=-90]{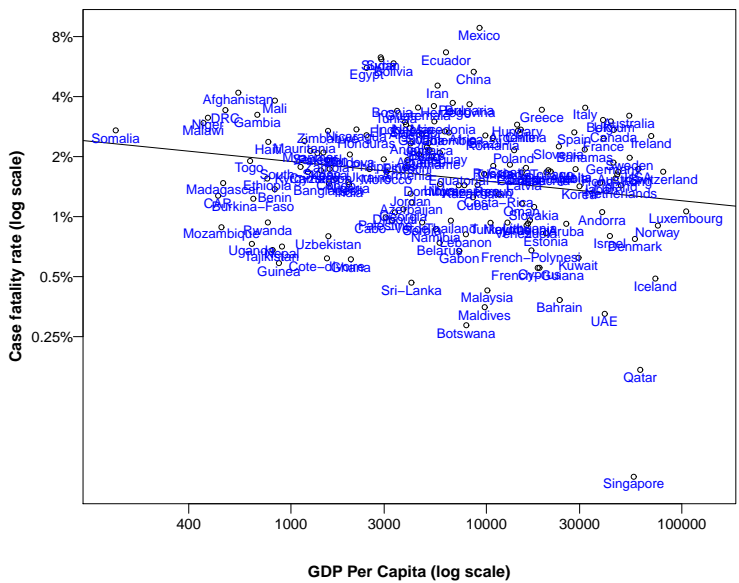}
\end{center}
\abovecaptionskip=-5pt
\caption{\it Scatter plot of CFR by GDP per capita for individual countries as of Dec 28, 2020. The solid line is the regression line. }  
\label{figure:cfrGDP1228}
\end{figure}

Similar to the analysis on data as of July 6, 2020 in Section~\ref{section:analysis1228}, we carry out analysis on data as 
of Dec 28, 2020, where the total number of reported cases is 81, 597, 946 (more than 7 times of the July data) 
with a total death toll of 1,779,448 (slightly more than 3 times of the July data).
An overall observation is that most countries have a reduced observed 
CFR than that by the July 6 data. This is consistent with a widely acknowledged view that the CFR gradually drops with the 
on-going of the pandemic after certain stage. For example, the observed CFR for the US is 5.56\%, 5.43\%, 4.14\%, 3.09\%, 
2.87\%, 2.70\%,  2.35\%, 1.87\% as of May 6, June 6, through Dec 6, 2020,  respectively. This could be due to various reasons: 
the population handles better and better after learning from early lessons, further mutations of the COVID-19 virus may have caused 
it to be less lethal over time, or simply because of the lack of enough testings in earlier stages (which in the analysis is assumed to be 
uniformly distributed across the age groups, but not over time). 
\\
\\
We start by considering the effect of age on the CFR, using model \eqref{eq:lmAge0706}.
However, the result was a little surprising, and the median age of the population barely plays a role in the linear regression
which finishes with an almost 0 $R^2$, i.e., 0.0004152, and the p-value associated with the F-test at 0.802. To get some sense on
why this is the case, we plot the observed CFR for individual countries in Figure~\ref{figure:cfrCompare}. To 
facilitate easy comparison, we also include the observed CFR for data as of July 6, 2020. Figure~\ref{figure:cfrCompare}
is quite revealing, and we see that most of the countries with a high CFR as of July 6, 2020 have seen
a sharp decrease in their CFRs by Dec 28, 2020, while the decrease is marginal (or even increase a little) for those countries 
with a previously low CFR. The decrease trend is most significant for countries with a relatively high median age. 
\\
\\
We then include the GDPs and consider the following model
\begin{equation}
\label{eq:lmGDP1228}
\log(CFR) = \beta_0 + \beta_1 \cdot X + \beta_2 \cdot GDP + \epsilon,
\end{equation}
where $X$ is the median age of a population. We code GDP to be 1 if it is smaller than \$10,000 per capita and 2 otherwise. 
This leads to a reported $R^2$ at 0.0780 (adjusted 0.0656), and a p-value of $2.366\times 10^{-3}$ on F-test. The fitted model 
parameters are
\begin{equation*}
\beta_0=-3.9261, ~ \beta_1=0.0178, ~\beta_2=-0.5453.
\end{equation*}
The GDP is statistically significant with a p-value $5.25\times10^{-4}$, but the age is not as significant with a p-value of 0.0325. 
Similarly, we have produced the diagnostics as before 
which suggest that the regression residuals have a roughly constant variance over the range of fitted values except with a moderate 
departure from normality. Linear regression using the original GDP leads to slighter lower $R^2$. 
The effect of GDP on CFR can be visualized from Figure~\ref{figure:cfrGDP1228}, higher GDP leads to a lower CFR. This is 
consistent with our understanding, as higher GDP typically implies better public health and medical facilities.

\subsection{Findings in comparing the two analysis}
\label{section:compAnalysis}
We have carried out analysis of the CFR with the same models for COVID-19 data taken at two different
time snapshots. Much has happened during the time, with a fast increasing and then slowing down pattern of 
the pandemic in different countries during the summer, followed by the general upward trend into the winter. 
It will be interesting to compare the results we have obtained. To facilitate our comparison, we summarize
our results in Table~\ref{table:paramsComp}.
\begin{table}[h]
\begin{center}
\begin{tabular}{r|r||l|l}
    \hline
\multicolumn{2}{c|} {\textbf{Model}}  & $\bm{2020/07/06}$   	& $\bm{2020/12/28}$ \\[2pt]
    \hline
Model I   &Age                  & 0.0516 (1.91e-5)***     &-1.63e-3 (0.802)  \\[1pt]
                    &$R^2$         &0.1726 (0.164)          &4.15e-4 (-6.16e-3)  \\[2pt]
                     &F-stat         &20.23 (1.91e-5)***.    &6.34e-2 (0.8019) \\[2pt]
    \hline
Model II   &GDP      & 7.29e-6 (0.146)   &-0.3309(6.06e-3)**   \\[1pt]
                    &$R^2$         &0.0217 (0.0116) &0.0491 (0.0428) \\[2pt]
                     &F-stat         &2.152 (0.1457) &8.752 (6.06e-3)**  \\[2pt]
\hline
Model III   &Age& 7.14e-2 (2.81e-6)***    &1.78e-2 (0.0325)*  \\[1pt]
                    &GDP         & -0.5537 (0.0284)* &-0.5453 (5.25e-4)***  \\[2pt]
                    &$R^2$         & 0.2132 (0.1968) &0.0780 (0.0656)  \\[2pt]
                     &F-stat         & 13.00 (1.01e-5)*** &6.299 (2.37e-3)**  \\[2pt]
\hline
\end{tabular}
\end{center}
\caption{\it Estimation under different models for data during the first wave and second wave.} \label{table:paramsComp}
\end{table}
\\
\\
One particularly interesting observation is the reversing roles played by the two population covariates---age 
and GDP. Age is a significant covariate in the July 6 data, but 
no longer as important in the Dec 28 data; GDP is not an important covariate in the July 6 data but becomes 
significant in the Dec 28 data. What causes this? Our interpretation is that, by July 6, 2020, 
most of the countries are still trying to understand the mechanism of COVID-19 and exploring and learning 
how to effectively deal with COVID-19, so the quality of
public health and abundance of medical facilities have not yet been reflected in the CFR; rather the more fundamental
factor, the age played a major role at this stage. As time goes by, both the public and health workers are gaining 
experiences in the handling and treating of COVID-19, so the quality of medical care has picked up and becomes a major 
factor in the CFR of a country; by this time, the age effect starts to shrink. Note that such a statement applies when 
we attempt to compare CFRs of many countries simultaneously. Can we claim that the age effect
is mostly disappearing after nearly a year since the start of the pandemic? This motivates our analysis
in Section~\ref{section:ageInvariance}.
\subsection{Invariance of the age effect in CFR}
\label{section:ageInvariance}
To answer the question posed in Section~\ref{section:compAnalysis}, we will look at CFR by age groups and by countries. 
This will help get rid of the country effect in CFR due to the difference in their population age structures, and also to standardize 
many other factors caused by differences among countries. For simplicity and constrained by the availability of the data 
(unfortunately, for most of the countries in the world, such statistics breakdown by age groups are not available), we 
will use the same 11 countries that we use to produce Figure~\ref{figure:cfrAgeCountries}
and Figure~\ref{figure:cfrAgeCountriesLog}. We will additionally analyze the CFR by age groups for these 11 countries using
data around Dec 28, 2020. 
\\
\\
We first carry out a simple linear regression on CFR (in log scale) versus age groups for the 11 countries involved. We treat 
each group in a country as an instance of data. As the ages are given as a range, we take the middle of the age groups, i.e., 
10, 25, 35, ..., 75, and 85, in linear regression. This leads to a fairly good fit to the linear model on the July 6 data, with the 
estimated coefficients as follows
\begin{equation*}
\hat{\beta}_0=-3.005648, ~\hat{\beta}_1=0.071463,
\end{equation*}
and a reported $R^2$ at 0.9102 (adjusted 0.8952) and p-value less than 2.34e-4 for the F-test. So the age effect is significant, and in 
particular, there is an exponential increase in CFR with the moving up through age groups. 
\\
\\
A similar regression analysis is carried out using data as of Dec 28, 2020, from the same 11 countries. 
The model fits the data well, with a reported $R^2$ at 0.9730 (adjusted 0.9685), and a p-value of 6.20e-6 on the F-test. 
The fitted intercept and slope are as follows
\begin{equation*}
\hat{\beta}_0=-2.907619, ~\hat{\beta}_1=0.070382,
\end{equation*}
which are surprisingly close to that on data as of July 6, 2020. So from data separated about half year apart, we see 
the same exponential age effect with almost the same exponential factor between age groups. This suggests that the
exponential age effect is invariant (or nearly) regardless of countries and time. Given that the 11 countries have a wide spectrum
of median ages, ranging from 27.1 to 47.3, and GDP per capita, ranging from \$6,120 to \$54,075 per year. We expect
such an invariance to widely hold across countries. 
\begin{figure}[h]
\centering
\begin{center}
\hspace{0cm}
\includegraphics[scale=0.5,clip,angle=-90]{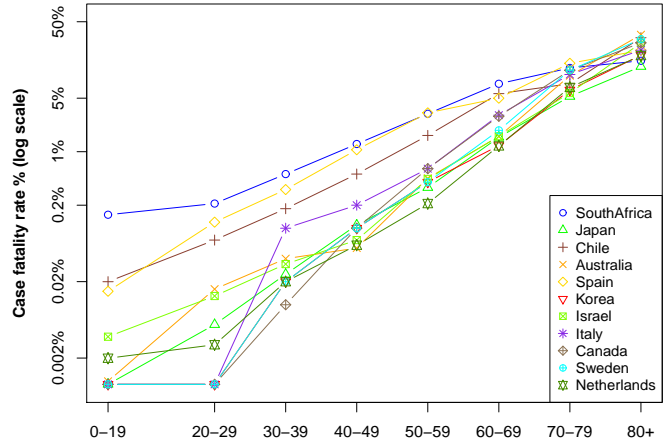}
\end{center}
\abovecaptionskip=-10pt
\caption{\it Log-scaled CFR by age groups for selected countries as of Dec 28, 2020.  }  
\label{figure:cfrAgeCountriesLog1228}
\end{figure}
\section{Conclusions}
\label{section:conclusion}
We have analyzed the CFR for countries in the world by including population covariates such as age and GDP, 
as proxy for the quality and abundance of healthcare. This allows us to understand the roles played by age 
and GDP in the apparently discrepant CFRs across countries despite the limitation of data accuracy. 
By analysis of data collected at two separate time snapshots, July 6 and Dec 28, 2020,
we have arrived at some interesting findings. During the initial stage of pandemic, age is a significant factor in 
CFR while GDP plays a less significant role, and then as the pandemic continues with the public and health 
workers gradually gaining experience in handling and treating COVID-19, GDP becomes a more significant
factor than the age. However, the exponential age effect is largely invariant 
across different age groups which is clearly exhibited on both data with nearly identical estimated 
exponent.
 
\section*{Appendix}
One aspect we omit in the main text is to consider the effect of different age groups to the CFR. To do this, we 
replace the median age in the linear model by the respective percentage of different age groups, namely, 20-29, 30-39, ..., 70-79, 
and 80+ in the population. The age group 0-19 is not included as the percentage of all age groups add up to 1. 
This leads to the following model
\begin{multline}
\log(CFR)=\beta_0+ \beta_2 \cdot X_{20-29} + \beta_3 \cdot X_{30-39} + \beta_4 \cdot X_{40-49} + \\
\beta_5 \cdot X_{50-59} + \beta_6 \cdot X_{60-69} + \beta_7 \cdot X_{70-79} + \beta_8 \cdot X_{80+} 
\label{model:lmAgeProf}
\end{multline}
where $X_{.}$'s are the percentage of respective age groups in a population. 
Again, we exclude countries with less than 3000 reported cases. On the July 6 data, the parameters fitted by linear 
regression are as follows
\begin{equation*}
\bm{\beta}=(-1.606, -8.799, -3.742, -3.016, -1.178, -12.905, 23.548, 3.171),
\end{equation*}
with a reported $R^2$ at 0.3993 (adjusted 0.3531) and a p-value at $4.271\times10^{-8}$ on F-test. 
\\
\\
What is interesting about Model~\ref{model:lmAgeProf} is that, the regression coefficients 
$\beta_{2-6}$ are all negative and $\beta_{7-8}$ are positive. The former implies that the increasing of respective variable 
value will lead to a decrease of CFR due to a higher proportion of younger people in the population, while the latter implies 
that the increasing of respective variable value will result in a larger CFR as there would be more senior people (i.e., age 70+) 
in the population. Additionally, the coefficients $\beta_{2-5}$ are increasing. While the actual value may be noisy, {\it qualitatively} 
this implies that, below 60, the younger age groups are more important in reducing the overall CFR. This is quite expected, and 
consistent with the exponential increasing trend of age-specific CFRs shown in Figure~\ref{figure:cfrAgeCountries}.
Two age groups that are particularly interesting are 60-69 and 70-79, which are playing opposite roles to the overall CFR. One 
possible interpretation might be that these two age groups lies at the age boundary just before and when the CFR quickly takes off. 
These two age groups have a major impact to the overall CFR. It may be worthwhile to allocate more resources to the particularly 
vulnerable age group 70-79 to reduce the overall CFR for a sizable population. The impact by the age group 80+ is less, 
which we attribute to its smaller percentage in the population.
Similar post-regression diagnostic analysis can be carried out, and we omit them here. 
\\
\\
A similar analysis can be carried out on the Dec 28 data, with estimated coefficients 
\begin{equation*}
\bm{\beta}=(-2.061, -6.335, -3.827, -2.017, -1.176, -5.163, 3.098, 1.606),
\end{equation*}
and a reported $R^2$ at 0.123 and a p-value of $7.52\times 10^{-3}$ on the F-test. Similar as for the July 6 data, all
the coefficients $\beta_{2-6}$ are negative and $\beta_{2-5}$ exhibit a decreasing trend when moving towards
a higher age group. Also observed is the similar special role by the age group 60-69 and 70-79. 
\\
\\
It is remarkable that by simply 
providing the observed CFR and the respective percentage of different age groups
for a number of countries, the data is actually able to speak about the desired age effect. 

\begin{thebibliography}{10}

\bibitem{ChakravartiLahaRoy1967}
I.~Chakravarti, R.~Laha, and J.~Roy.
\newblock {\em {Handbook of Methods of Applied Statistics, Volume I}}.
\newblock John Wiley and Sons, 1967.

\bibitem{CookWeisberg1983}
R.~D. Cook and S.~Weisberg.
\newblock Diagnostics for heteroscedasticity in regression.
\newblock {\em Biometrika}, 70(1):1--10, 1983.

\bibitem{GuptaShankar2020}
S.~Gupta and R.~Shankar.
\newblock {Estimating the number of COVID-19 infections in Indian hot-spots
  using fatality data}.
\newblock {\em arXiv:2004.04025}, 2020.

\bibitem{JagodnikLachmann2020}
K.~M. Jagodnik, F.~Ray, F.~M. Giorgi, and A.~Lachmann.
\newblock {Correcting under-reported COVID-19 case numbers: estimating the true
  scale of the pandemic}.
\newblock {\em medRxiv doi 10.1101/2020.03.14.20036178}, 2020.

\bibitem{UNPopulation}
United Nations.
\newblock {\em World Population Prospects}, 2019.
\newblock {https://population.un.org/wpp/Download/Standard/Population/}.

\bibitem{PhilipRaySubramanian2020}
M.~Philip, D.~Ray, and S.~Subramanian.
\newblock {Decoding India's Low Covid-19 Case Fatality rate}.
\newblock {\em Working Paper 27696, Natioanal Bureau of Economic Research,
  http://www.nber.org/papers/w27696}, 2020.

\bibitem{Rice1995}
J.~A. Rice.
\newblock {\em Mathematical Statistics and Data Analysis}.
\newblock Duxbury Press, 1995.

\bibitem{Schroder2020}
I.~Schr${\ddot{\text{o}}}$der.
\newblock {COVID-19}: A risk assessment perspective.
\newblock {\em ACS Chemical Health and Safety}, 27(3):160--169, 2020.

\bibitem{ShimMizumotoChoi2020}
E.~Shim, K.~Mizumoto, W.~Choi, and G.~Chowell.
\newblock {Estimating the risk of COVID-19 death during the course of the
  outbreak in Korea, February-May, 2020}.
\newblock {\em Journal of Clinical Medicine}, 9(6):1641, 2020.

\bibitem{KugelgenGreseleSchkolpf2020}
J.~von K${\ddot{\text{u}}}$gelgen, L.~Gresele, and
  B.~Sch${\ddot{\text{o}}}$lkopf.
\newblock {Simpson’s paradox in Covid-19 case fatality rates:a mediation
  analysis of age-related causal effects}.
\newblock {\em arXiv:1604.00989}, 2016.

\bibitem{WikiWorldMedianAge}
Wikipedia.
\newblock {\em {List of countries by median age}}, 2020.
\newblock {https://en.wikipedia.org/wiki/List\_of\_countries\_by\_median\_age}.

\bibitem{covidWorldOmeters}
Worldometer.
\newblock {\em {COVID-19 Coronavirous Pandemic}}, 2020.
\newblock {https://www.worldometers.info/coronavirus/}.

\bibitem{YanXuWang2020}
D.~Yan, Y.~Xu, and P.~Wang.
\newblock Estimating the number of infected cases in {COVID-19} pandemic.
\newblock {\em Journal of Data Science (in press), arXiv:2005.12993}, 2020.

\end{thebibliography}

\end{document}